\documentclass[12pt]{article}
\usepackage{ amsmath, amsthm, amssymb } 
\usepackage{adjustbox}
\usepackage{mathtools,slashed}

\usepackage{dsfont}  

\usepackage{url}  
\usepackage{color} 
\usepackage{cite}

\newcommand{\be} { \begin{equation} } 
\newcommand{\ee} { \end{equation} } 
\newcommand{\I}{\mathcal{I}}

\newcommand{\fin}{{\rm fin}}

\newcommand{\uv}{\rm UV}
 
\newcommand{\p}{p\hspace{-0.9ex}/}

\newcommand{\un}{{\rm un}}

\newcommand{\ep}{\epsilon}

\begin{document} 
\allowdisplaybreaks

\setlength{\unitlength}{1.3cm} 
\begin{titlepage}
\vspace*{-1cm}
\begin{flushright}
TTP17-055
\end{flushright}                                
\vskip 3.5cm
\begin{center}
\boldmath

{\Large\bf Two-loop amplitudes for processes $g g \to H g, q g \to H q$ and $q \bar{q} \to H g$ 
at large Higgs transverse momentum \\[3mm] }
\unboldmath
\vskip 1.cm
{\large , Kirill Kudashkin}$^{a,}$
\footnote{{\tt e-mail: kirill.kudashkin@kit.edu}} 
{\large Kirill Melnikov}$^{a,}$
\footnote{{\tt e-mail: kirill.melnikov@kit.edu}}
{\large , Christopher Wever}$^{a,b,}$
\footnote{{\tt e-mail: christopher.wever@kit.edu}} 
\vskip .7cm
{\it $^a$  Institute for Theoretical Particle Physics, KIT, 76128 Karlsruhe, Germany} \\
{\it $^b$  Institut f\"ur Kernphysik, KIT, 76344 Eggenstein-Leopoldshafen, Germany} 
\end{center}
\vskip 2.6cm

\begin{abstract}
\noindent
We compute the two-loop QCD corrections to  amplitudes for processes 
$g g \to H g, q g \to H q$ and $q \bar{q} \to H g$ in the 
limit when the Higgs transverse momentum is larger  than the 
top quark mass, $p_\perp \gg m_t$.   These amplitudes are important ingredients for understanding 
higher-order QCD effects on Higgs transverse momentum distribution at large $p_\perp$. 

\vskip .7cm 
{\it Key words}: QCD, Higgs physics, multi-loop computations, asymptotic expansion, large transverse momentum
\end{abstract}
\vfill
\end{titlepage}                                                                
\newpage

\section{Introduction} \label{sec:intro} \setcounter{equation}{0} \setcounter{footnote}{0}  
\numberwithin{equation}{section}

The discovery of the  Standard Model (SM)-like Higgs boson at the LHC five years ago 
got rapidly  transformed into an active 
experimental  exploration of this new particle. Indeed, a detailed  knowledge of  Higgs boson properties 
and its coupling to other particles is essential for understanding its role 
in the electroweak symmetry breaking and for early clues about physics beyond the Standard Model. 
Since in the SM the Higgs couplings to gauge bosons and matter particles  
can  be computed theoretically to a very high precision, the existence of equally precise 
measurement program is crucial to  search for differences between measurements and predictions 
that may then be interpreted as signals of physics beyond 
the Standard Model (BSM).

Unfortunately, most recent results from the Run II of the LHC show that 
the Higgs boson fits very well the expected profile of the  SM Higgs particle 
 and no signs of New Physics 
have been seen so far. These conclusions  are so far limited by 
statistical and systematic errors that, on average, are in the ${\cal O}(15-20)$ percent  range but 
can be much  larger for certain couplings and cross sections. 
It is expected that during the  Run II  and the high-luminosity phase of the LHC,  
the precision of Higgs couplings measurements will significantly increase, 
reaching eventually a  few percent accuracy. 

This accuracy has to be matched on the theory side and we have seen quite very impressive 
accomplishments in refining predictions for major Higgs production and decay processes in recent 
years. For example, the inclusive Higgs boson production in gluon fusion 
is now known to an impressive next-to-next-to-next-to-leading order (N$^3$LO) QCD in 
the infinite top quark mass limit~\cite{Anastasiou:2016cez}  and the $H+$jets cross section 
has been computed through next-to-next-to-leading order (NNLO) QCD in the same 
approximation~\cite{Boughezal:2013uia,Chen:2014gva,Boughezal:2015dra,Boughezal:2015aha}. 

The approximation of an infinitely heavy top quark is justified as long as  typical 
values of kinematic parameters relevant for  particular cross sections are smaller than ${\cal O}(2 m_t)$. 
Although this criterion is satisfied for the majority of events selected for both  inclusive and 
$H+j$ cross sections, there are good reasons to look at regions of phase-space where this condition 
is explicitly violated.  For example,  with the dramatic  increase of 
statistics promised by  the high-luminosity run at  the LHC, we will have access to 
Higgs transverse momentum distribution at high values of $p_\perp \ge m_t$. This is a very interesting 
regime since, as a matter of principle,  it allows us to disentangle two terms  in the effective SM 
Lagrangian -- the point-like Higgs coupling to gluons   and the modification of the Higgs-top 
Yukawa coupling~\cite{Arnesen:2008fb,Harlander:2013oja,Azatov:2013xha,Grojean:2013nya,Banfi:2013yoa}.\footnote{See~\cite{Neumann:2016dny}  for further references.}  
Amazingly,  first experimental attempts to explore Higgs boson production  
at high-$p_\perp$  have recently  been undertaken~\cite{CMS:2017cbv}.


To fully benefit from this opportunity, it is important 
to have as precise predictions for Higgs $p_\perp$-distribution at large transverse momenta as 
possible.   Since  for computations  at high $p_\perp \ge m_t$, 
the Higgs coupling to gluons cannot be treated  as point-like, all existing 
higher-order computations, including most recent NNLO QCD predictions for 
$H+j$ production~\cite{Boughezal:2013uia,Chen:2014gva,Boughezal:2015dra,Boughezal:2015aha}
are of little use.  In fact,  when mass effects are accounted for, 
the $p_\perp$-distribution appears to be  known only at  leading order which, in this case, 
is determined by one-loop diagrams.  Since NLO QCD corrections for processes with 
gluons in initial state are known  to be large \cite{Dawson:1990zj,Djouadi:1991tka,Spira:1995rr}, it is quite conceivable 
that large corrections to Higgs transverse momentum distribution at high $p_\perp$ 
are to be found as well.  Computing two-loop contributions to relevant amplitudes 
and setting up the stage for a full NLO QCD computation of the Higgs boson transverse 
momentum distribution at high $p_\perp$ is the main goal of this paper. 

We note that the 
relevant two-loop amplitudes for a NLO computation of Higgs plus jet production mediated 
via a massive quark-loop were considered recently  
in ~Refs.\cite{Melnikov:2016qoc,Melnikov:2017pgf}. However, 
in those papers the  limit of a small quark mass $m_q \ll m_H \sim  p_\perp$  was considered.  
This limit is relevant for the bottom quark contribution to effective $ggH$ interaction vertex but 
it is not the right limit to describe  high-$p_\perp$ regime of the Higgs boson production.

To address the high-$p_\perp$ case  we impose the following 
hierarchy between kinematic variables and particle masses 
$m_h^2\ll m_t^2\ll s,t,u$. This result is then applicable to the case where the Higgs boson 
is produced  via a top quark loop at high $p_\perp$.\footnote{We consider all quarks beyond 
the top quark   to be massless. The 
contribution of the bottom-quark loop has been considered in~\cite{Lindert:2017pky} and was 
found to be  negligible.} 
To compute the scattering amplitude in that limit, we will follow an approach 
developed in~Refs.\cite{Mueller:2015lrx,Melnikov:2016qoc,Melnikov:2017pgf} 
and expand the relevant Feynman integrals in  small parameters, namely in 
$m_h^2/m_t^2$ and $m_t^2/s$, using the  differential equations that these Feynman integrals satisfy. 
We note that the computation of relevant  integrals for arbitrary Higgs and quark masses 
is still ongoing; planar  master integrals have recently  been computed in~\cite{Bonciani:2016qxi}. 

The remainder of the paper is organized as follows.  In Section~\ref{sec:notation} 
we explain the notation, introduce the relevant amplitudes, explain their decomposition 
into invariant form factors 
and describe the renormalization. 
In Section~\ref{sec:formfactors} we discuss  how form factors 
are computed. 
 We explain how to calculate the master integrals with the differential equation method  
in Section~\ref{sec:masters}. In Section~\ref{sec:bound} we provide  
an example of how  integration constants for differential equations can be computed. 
The final results for helicity amplitudes are presented in Section~\ref{sec:helamp}. The 
amplitudes are originally computed in the kinematic region where $t>0,s,u<0$; 
in Section~\ref{sec:cont} we describe the analytic continuation to other relevant scattering regions. 
We conclude in Section~\ref{sec:conclusions}.  
We include  ancillary files with this submission  that 
contain  analytic results for all relevant amplitudes in the different kinematic regions.

\section{The scattering amplitudes}
\label{sec:notation} \setcounter{equation}{0} 
\numberwithin{equation}{section}

\begin{figure}[t!]
\centering
\includegraphics[width=0.25 \linewidth]{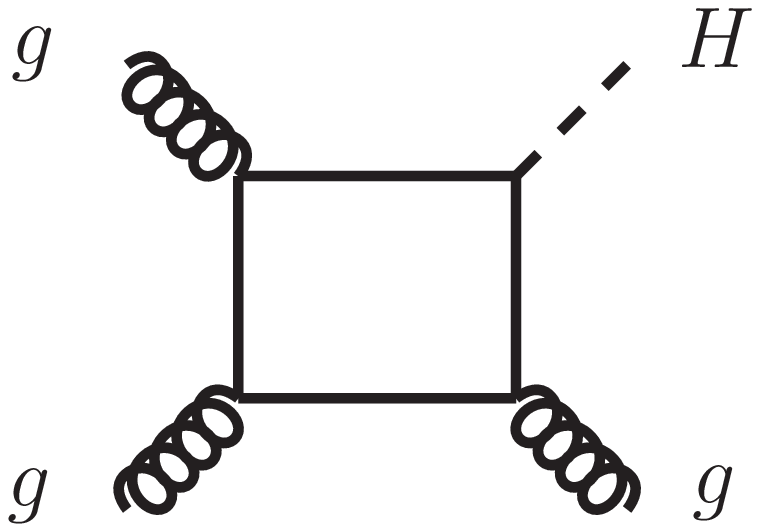} \hspace{0.4 cm}
\includegraphics[width=0.25 \linewidth]{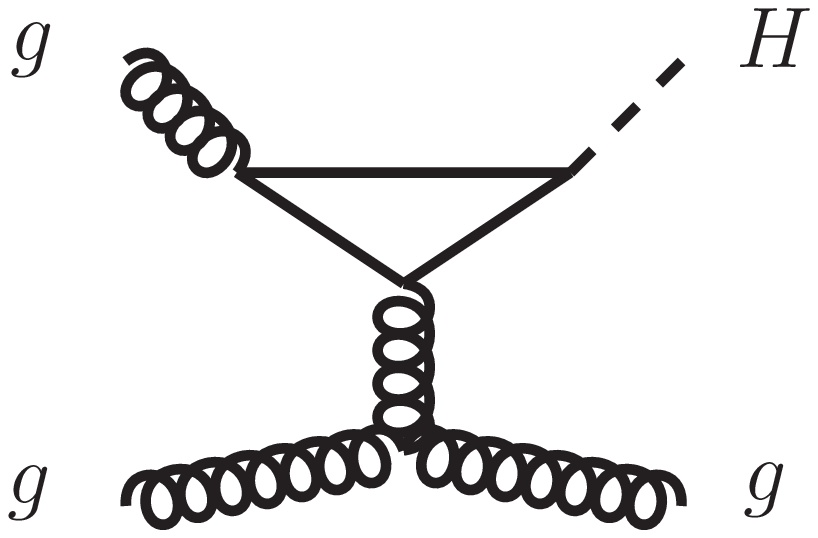} \hspace{0.4 cm}
\includegraphics[width=0.25 \linewidth]{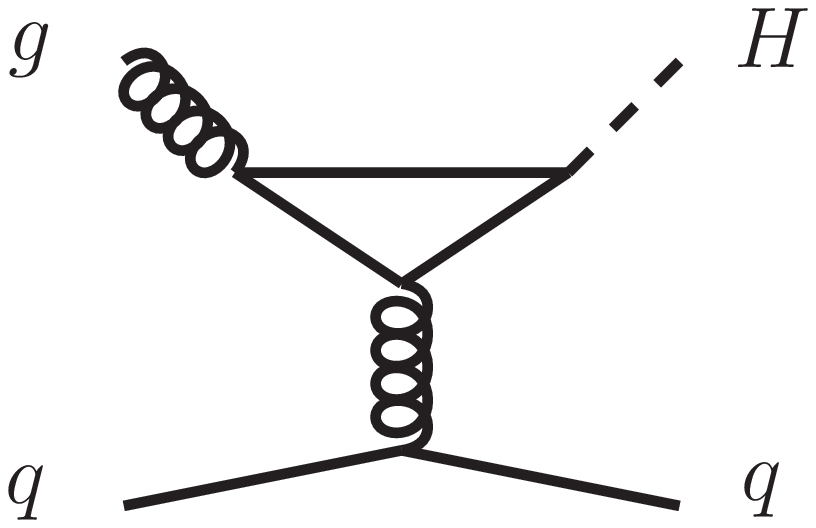} 
\caption{The one-loop Feynman diagrams that contribute to the quark-loop induced processes 
$gg\rightarrow Hg$ and $qg\rightarrow Hq$.}
\label{fig::feyndiag1loop}
\end{figure}

Production of the Higgs boson in association with  a jet  at a hadron collider can occur in several 
different ways;   the relevant partonic processes can be found by crossing the Higgs decay processes
\begin{gather}
H(p_4) \to g(p_1) + g(p_2) + g(p_3), \nonumber\\
H(p_4) \to q(p_1) + \bar{q}(p_2) + g(p_3),\,
\label{eq:Htoqqg}
\end{gather}
to the production kinematics.  We consider all quarks in Eq.(\ref{eq:Htoqqg}) as massless. 
The Higgs boson interaction with gluons and massless quarks is facilitated by loops of top quarks; 
this is the only quark that we consider massive in this article. 
Some examples of Feynman diagrams that contribute to (crossed versions of)  processes shown 
in Eq.(\ref{eq:Htoqqg})  are presented 
in Fig.~\ref{fig::feyndiag1loop}.  The goal of this 
paper is to compute   two-loop contributions to scattering amplitudes for 
processes in Eq.(\ref{eq:Htoqqg}) assuming that the Higgs boson mass and the top quark mass 
are smaller than all other kinematic invariants. 

We start by defining the Mandelstam variables
\be
s = (p_1 + p_2)^2\,, \quad t = (p_1 + p_3)^2\,, \quad u = (p_2 + p_3)^2\,,\quad
s + t + u = m_h^2.
\ee
We trade four  dimensionful Mandelstam variables for a dimensionful variable 
$s$ and  three dimensionless variables
\begin{equation}
\eta = -\frac{m_h^2}{4m_t^2}\,, \quad \kappa = - \frac{m_t^2}{s}\,, \quad z = \frac{u}{s}\,. \label{eq:variables}
\end{equation}
In the large transverse momentum region and in the limit of a small Higgs mass the following hierarchy of scales applies 
\begin{equation}
m_h^2, m_t^2 \ll |s| \sim  |t| \sim |u| \quad \rightarrow \quad |\eta|, |\kappa| \ll 1, \, |z| \sim 1.
\label{eq:hierarchy}
\end{equation}
For the top quark and Higgs boson with masses $m_t\sim 173$ GeV and $m_h\sim 125$ GeV respectively, 
$|\eta | \sim 0.13$ and can be treated as a small parameter.

A Euclidean region where all Mandelstam variables $s,t$ or $u$ are negative does not exist since 
$|m_h^2| = |s+t+u|\ll |s|,|t|,|u|$ in the kinematic region that is of interest to us. 
At least one of the Mandelstam 
variables has to be 
positive and without loss of generality we choose $t$ to be positive and $s,u$ negative. Furthermore we will compute 
our amplitudes initially in the region where $m_h^2<0$ and $m_t^2>0$, in other words the parameters will satisfy
\begin{equation}
0<\eta \ll 1\,, \quad 0<\kappa \ll 1\,, \quad 0<z\,, \;\;\; \,s < 0\, \label{eq:region}.
\end{equation}
If we analytically continue  to the region where 
$m_h^2>0$, our results will represent the physical scattering processes
\begin{gather}
g(-p_1) + g(-p_3) \to H(-p_4) + g(p_2)\, \nonumber\\
\bar{q}(-p_1) + g(-p_3) \to H(-p_4) + \bar{q}(p_2)\,.
\label{eq:scattering}
\end{gather} 
All other production channels can be found from crossing and analytic continuation of the computed amplitudes in 
the region specified in Eq.~\eqref{eq:region}, as we 
will describe in Section~\ref{sec:helamp}. Note that because the 
Euclidean region does not exist, all the amplitudes 
have explicit imaginary parts. 

We follow ~Refs.\cite{Melnikov:2016qoc, Melnikov:2017pgf} and define the partonic amplitudes 
corresponding to the processes shown in Eq.~\eqref{eq:Htoqqg} as
\begin{align}
& {\cal A}_{H\rightarrow ggg}\left (p_1^{a_1},p_2^{a_2},p_3^{a_3} \right )  = 
f^{a_1 a_2 a_3}\;  \epsilon_1^\mu \,\epsilon_2^\nu \, \epsilon_3^\rho \, {\cal A}^{g}_{\mu \nu \rho}(s,t,u, m_t)\,, \\
& \mathcal{A}_{H\rightarrow q\bar{q}g}(p_1^j, p_2^k, p_3^a) = i\, T^{a}_{jk}\;
\epsilon^\mu_3(p_3)\, \bar{u}(p_1)\, \mathcal{A}^{q}_\mu(s,t,u,m_t)\, v(p_2) \,.
\label{eq:ampl}
\end{align}
The color structure of the amplitudes is completely factorized and captured by the $SU(3)$ structure 
constants $f^{a_1 a_2 a_3}$ and the usual Gell-Mann matrices $T^{a}_{jk}$ for the 
gluon and quark channels respectively. 
The color indices are denoted by $a_{1,2,3}$ and $j,k$ for gluons and quarks, respectively.
The gluon polarization vectors are transversal $\epsilon_i \cdot p_i = 0$, $i=1,2,3$ and 
the  spinors satisfy the massless Dirac equations $\slashed{p}_1u(p_1)=\slashed{p}_2v(p_2)=0$.

To understand the Lorentz structure of the amplitude,  we write it  as a sum of 
parity conserving Lorentz tensors of  relevant ranks.  
The amplitudes must furthermore satisfy the Ward identity which 
implies that an on-shell amplitude must vanish  after replacing any of the gluon polarization vectors 
with  their momenta. After imposing these constraints, the $H\rightarrow ggg$ and 
$H\rightarrow q\bar{q}g$ amplitudes can be written as a sum of four (two) tensors, 
respectively. They read  
\be 
\begin{split} 
{\cal A}^{g}_{\mu \nu \rho}(s,t,u, m_t) &= F^{g}_1\, g_{\mu \nu}\, p_{2\rho}
 + F^{g}_2\, g_{\mu \rho}\, p_{1\nu} 
+ F^{g}_3\, g_{\nu \rho}\, p_{3\mu}
 + F^{g}_4\, p_{3\mu} p_{1\nu} p_{2\rho}\,, \\
\mathcal{A}^{q}_\mu &= F^{q}_1\, \left( \p_3\, p_{2\mu} - p_2 \cdot p_3  \, \gamma^\mu \right)
+ F^{q}_2 \left( \p_3\, p_{1\mu} - p_1 \cdot p_3  \, \gamma_{\mu} \right ). 
\label{eq:tensor}
\end{split}
\ee
The above decomposition corresponds to the  
cyclic gauge fixing condition for the gluon polarization  vectors
\be
\epsilon_1 \cdot p_2 = \epsilon_2 \cdot p_3 = \epsilon_3 \cdot p_1 = 0. \label{eq:gauge}
\ee
The form factors $F^{q,g}_j$  are scalar functions 
of the Mandelstam variables  and the quark mass. In the following we will drop 
the upper index ${q}$ and ${g}$ for simplicity, unless 
they need to be  explicitly specified.

\begin{figure}[t!]
\centering
\includegraphics[width=0.25 \linewidth]{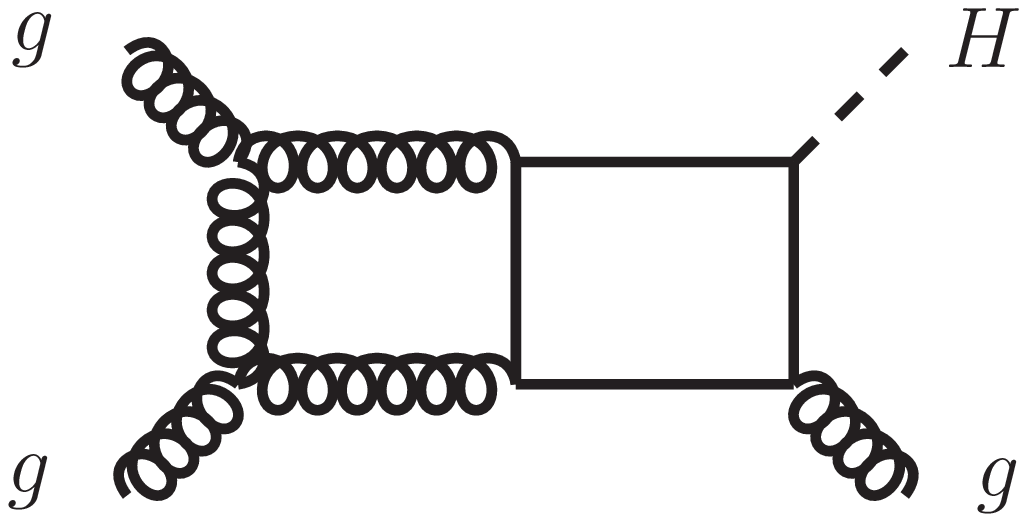} \hspace{0.4 cm}
\includegraphics[width=0.25 \linewidth]{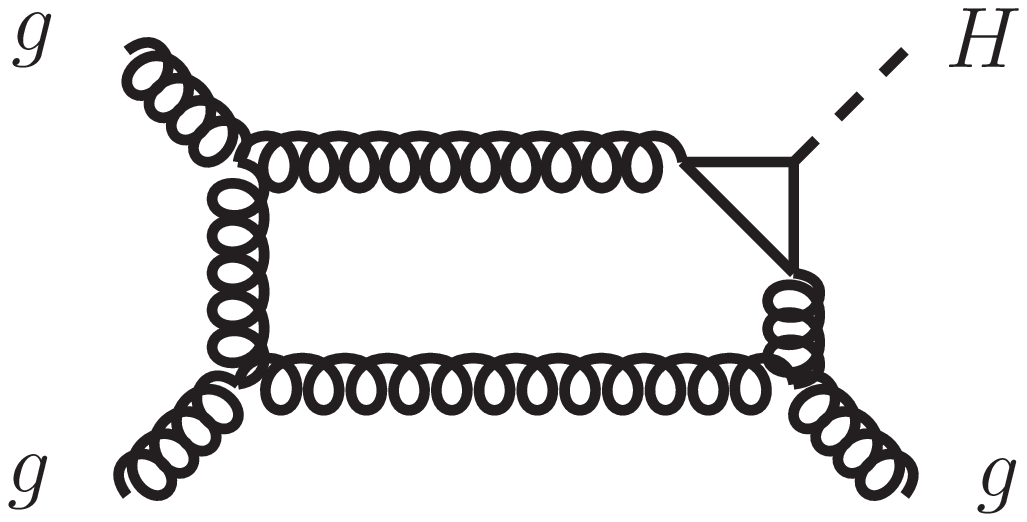} \hspace{0.4 cm}
\includegraphics[width=0.25 \linewidth]{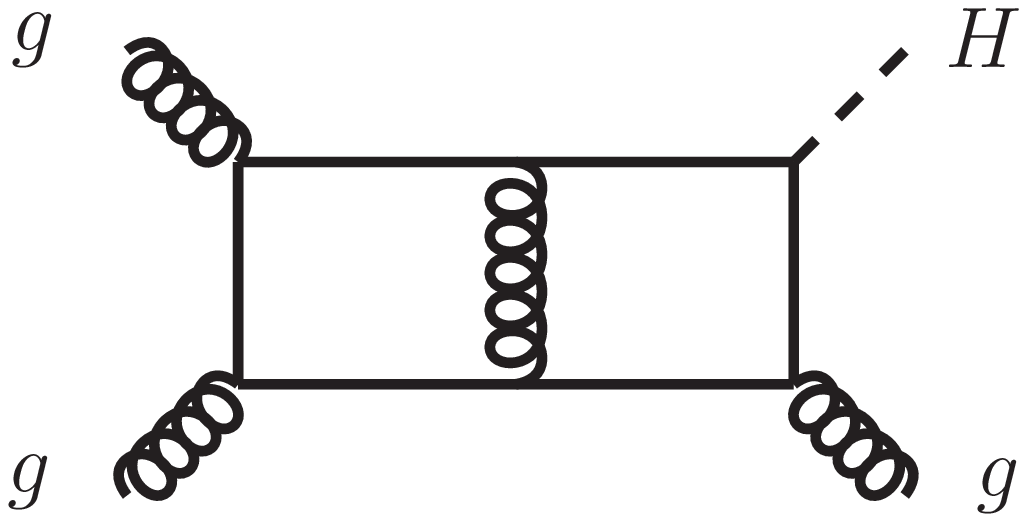} \hspace{0.4 cm}\\
\includegraphics[width=0.25 \linewidth]{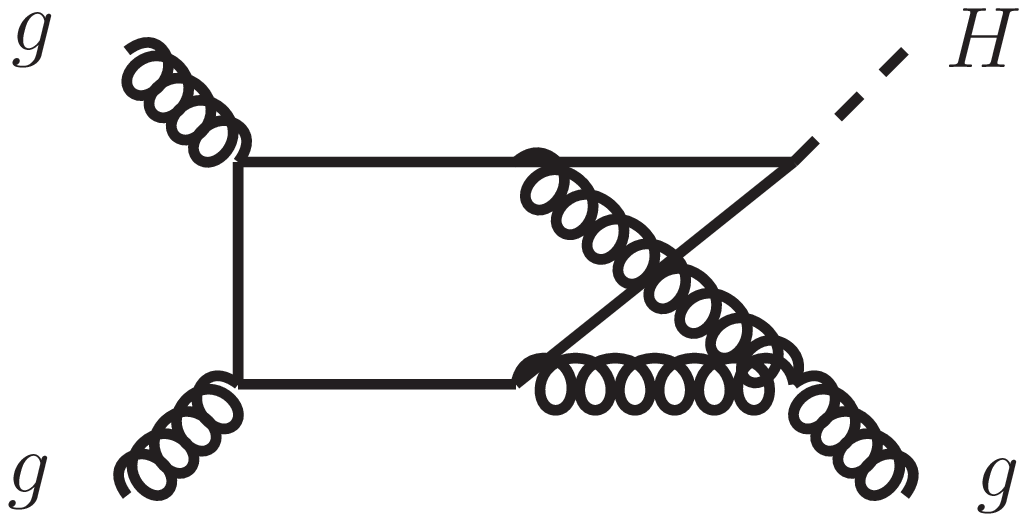} \hspace{0.4 cm}
\includegraphics[width=0.25 \linewidth]{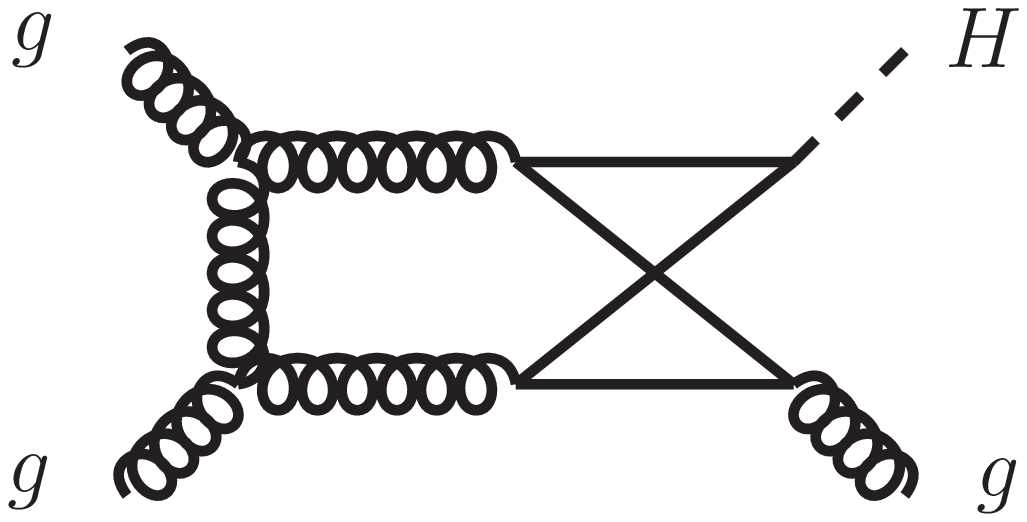}
\caption{Examples of two-loop Feynman diagrams that contribute to the process $gg\rightarrow Hg$.}
\label{fig::feyndiag2loopgg}
\end{figure}

\begin{figure}[t!]
\centering
\hspace{-2 cm} \includegraphics[width=0.35 \linewidth]{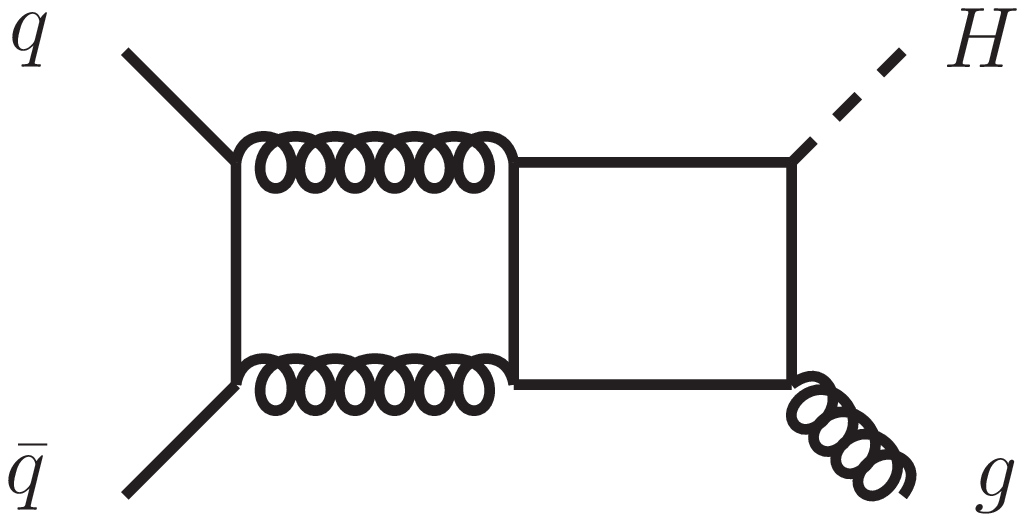} \hspace{-0.4 cm}
\includegraphics[width=0.35 \linewidth]{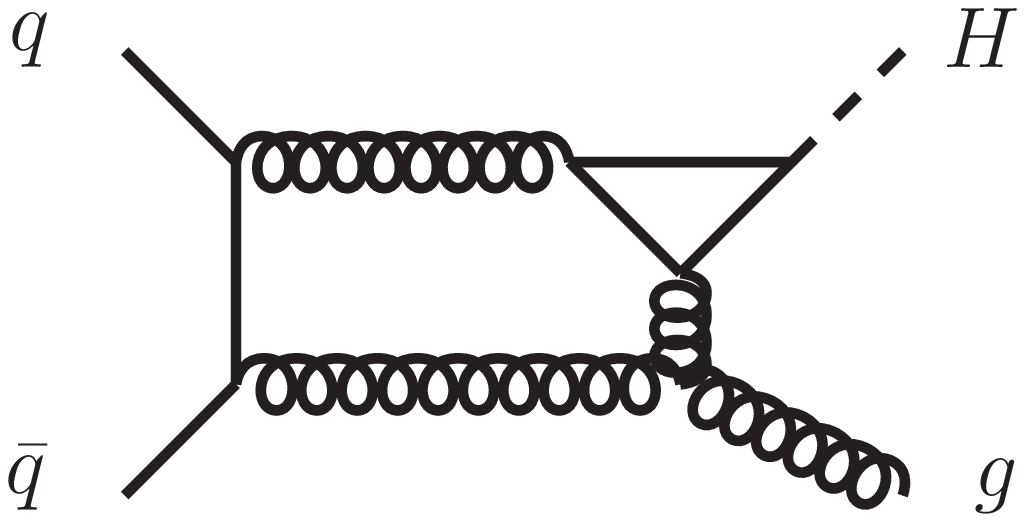} \hspace{1.4 cm}
\includegraphics[width=0.25 \linewidth]{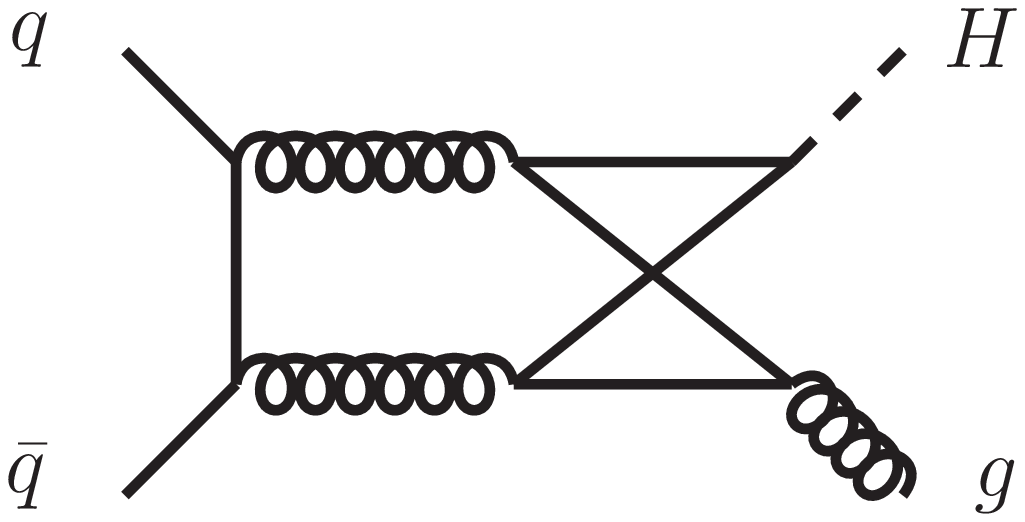}
\caption{Examples of two-loop Feynman diagrams that contribute to the process $q\bar{q}\rightarrow Hg$.}
\label{fig::feyndiag2loopqq}
\end{figure}

The unrenormalized form factors $F_j$ can be expanded in the bare 
QCD coupling constant constant $\alpha_0$ as
\begin{equation}
F_j^{\un}(s,t,u,m_t) = \sqrt{\frac{\alpha_0^3}{\pi}} \left[ F_j^{(1),\un} 
+ \left(\frac{\alpha_0}{2 \pi}\right) F_j^{(2),\un}  + \mathcal{O}(\alpha_0^2)\right]\,.
\label{eq:bareF}
\end{equation}
The LO contribution $F_j^{(1)}$ with the full dependence on the quark mass 
was calculated in Refs.~\cite{Ellis:1987xu,Baur:1989cm}. In this paper,  we will compute 
the two-loop contributions to form factors  $F_j^{(2)}$ 
assuming that the Higgs boson transverse momentum is large and the Higgs boson mass is parametrically 
smaller than the mass of the top quark.  Some examples of two-loop diagrams  that contribute 
to Higgs boson production in association with a jet are shown 
in Figs.~\ref{fig::feyndiag2loopgg} and~\ref{fig::feyndiag2loopqq}. 

The unrenormalized form factors that we compute have poles in $\epsilon=(d-4)/2$; these poles are of 
ultraviolet (UV) and/or infrared (IR) origin. We perform the subtraction of these poles in two steps. First we 
UV renormalize the above bare form factors $F_j^{\un}$ in Eq.~\eqref{eq:bareF}
\begin{equation}
F_j^{\uv}(s,t,u,m_t) = \sqrt{\frac{ \alpha_s^3}{\pi\, S_\epsilon^3}}\,  
\left[  F_j^{(1),\uv}
+  \left( \frac{\alpha_s}{2 \pi}  \right) F_j^{(2),\uv}  + \mathcal{O}(\alpha_s^3) \right]\,.
\end{equation}
We express the bare strong coupling 
constant and the top quark mass parameter in $F_j^{\un}$ in terms of renormalized parameters 
and we include for each external gluon the wave-function renormalization factor. The strong coupling constant 
gets renormalized in the mixed scheme; this implies that contributions of $N_f$ massless quarks 
are renormalized in the  $\overline{\text{MS}}$-scheme whereas 
top  quark contributions are subtracted at zero momentum.
The top quark  mass is renormalized in an  on-shell scheme. 
The corresponding formulas read 
\begin{gather}
\alpha_0\, \mu_0^{2 \epsilon} \; S_\epsilon = \alpha_s\, \mu_R^{2 \epsilon}\, 
\left[ 1 - \frac{1}{\epsilon} \left( \beta_0 
+ \delta_w  \right ) \; \left( \frac{\alpha_s}{2 \pi}  \right) + \mathcal{O}(\alpha_s^2)\right], \\
m_{t,0} = m_t\left[ 1 +  \left( \frac{\alpha_s}{2 \pi}  \right)\, \delta_m + \mathcal{O}(\alpha_s^2)\; \right]\,.
\label{eq:baren}
\end{gather}
Here $S_\epsilon = (4 \pi)^\epsilon\, e^{-\epsilon \, \gamma_E}\,, \;\; \gamma_E = 0.5772..$,
$\beta_0 = 11/6\; C_A - 2/3\, T_R \, N_f$, $T_R = 1/2$ and $C_A=N_c$ is the number of colors. The wave-function and 
mass renormalization constants are 
\be
\delta_w =  -2/3\;T_R (m_t^2/\mu_R^2)^{-\epsilon}, \quad \delta_m = C_F
\left( \frac{m_t^2}{\mu_R^2}\right)^{-\epsilon} 
 \left(  -\frac{3}{2\epsilon} -2 + \mathcal{O}(\epsilon) \right).\, 
\ee

Renormalization of the gluon wave-function is 
taken into account by multiplying the form factors with 
$$
\sqrt{Z_A} = 
1 + \frac{1}{2} \left( \frac{\alpha_s}{2 \pi}  \right)\,
\delta_w  + \mathcal{O}(\alpha_s^2), 
$$
for each of the external gluons. 

Following the described procedure, we express the UV-renormalized form factors 
in terms of bare ones. We find 
\begin{align}
 (F^{i})_j^{(1),\uv} &=  (F^{i})_j^{(1),\un}\,, \nonumber \\ 
 (F^{i})_j^{(2),\uv} &= S_\epsilon^{-1}\, (F^{i})_j^{(2),\un}  
 - \left(\frac{3 \,\beta_0}{2\,\epsilon} + \delta_{i,q}\frac{\delta_w}{\epsilon}\right)\; (F^{i})_j^{(1),\un} 
 + \, m_t\; \frac{d(F^{i})_j^{(1),\un}}{d m_t}\, \delta_m
 \,.
 \label{eq:UVren}
\end{align}
where $i=q,g$ denotes the $H\rightarrow ggg$ and $H\rightarrow q\bar{q}g$ form factors respectively.

Unfortunately, even after the UV renormalization is performed, the form factors still exhibit 
poles in $\epsilon$. These are the infra-red and collinear poles that appear in the virtual 
amplitude; they disappear once elastic and inelastic partonic processes are combined to 
compute physical cross sections.  Since the structure of IR-singularities is 
universal~\cite{Catani:1998bh} and since they, as we said,  will eventually get 
cancelled against real emission corrections, 
it is useful to separate them in the two-loop amplitude.  We write 
\be
(F^{i})_j^{(1), \uv} = (F^{i})_j^{(1), \fin} =  \,,\;\;\;\;\;
(F^{i})_j^{(2), \uv} = I^{i}_1(\epsilon)  (F^{i})_j^{(1),\uv} + (F^{i})_j^{(2),\fin},
\label{eq:IRsub}
\ee
where again $i=q,g$ and $I^{q,g}_1(\epsilon)$ are the so-called Catani operators
\begin{align}
I^{g}_1(\epsilon) =-\frac{ C_A e^{\epsilon \gamma}}{2 \Gamma(1-\epsilon)} &
\left( \frac{1}{\epsilon^2} + \frac{\beta_0}{C_A} \frac{1}{\epsilon}\right)
\left( \left(-\frac{s}{\mu_R^2}\right)^{-\epsilon} + \left(-\frac{t}{\mu_R^2}\right)^{-\epsilon}
 + \left(-\frac{u}{\mu_R^2}\right)^{-\epsilon} \right)\,, \\
I^{q}_1(\epsilon) = -\frac{ e^{\epsilon \gamma}}{2 \Gamma(1-\epsilon)} & 
\left( C_A\left( \frac{1}{\epsilon^2} + \frac{3}{4\,\epsilon} + \frac{\beta_0}{2\,C_A\,\epsilon}\right)\; \left(\left(-\frac{t}{\mu_R^2}\right)^{-\epsilon} + \left(-\frac{u}{\mu_R^2}\right)^{-\epsilon}\right) \right. \nonumber\\
& \left. -\frac{1}{C_A}\left( \frac{1}{\epsilon^2} + \frac{3}{2\,\epsilon}\right)\; 
\left(-\frac{s}{\mu_R^2}\right)^{-\epsilon} \right) .
\label{eq:cataniI1}
\end{align}
Our final results for the form factors, $\{F_j^{\rm fin}\}$ are finite in the limit $\epsilon \to 0$.
Note that in order to perform the final IR subtraction we require the one-loop amplitudes to order $\epsilon^2$.

\section{Computing the form factors} 
\label{sec:formfactors} \setcounter{equation}{0} 
\numberwithin{equation}{section}

The bare form factors are expressed in terms of Feynman diagrams that we produce with QGRAF~\cite{Nogueira:1991ex} 
and independently with FeynArts~\cite{Hahn:2000kx}. We allow for massless external quarks 
and both massive and massless internal quark loops. Some examples 
of Feynman diagrams that one has to consider are shown in 
Figs.~\ref{fig::feyndiag2loopgg} and~\ref{fig::feyndiag2loopqq}.
We follow procedures outlined in Refs.~\cite{Melnikov:2016qoc,Melnikov:2017pgf} 
to express the form factors in terms of scalar integrals by applying projection 
operators as follows
\be
\begin{split}
F^{g}_i(s,t,u,m_t) &= \sum_{pol}\, \mathcal{P}^{gi}_{\mu \nu \rho}\;
\epsilon_{1,\lambda_1}^{\mu,*} \epsilon_{1,\lambda_1}^{\mu_1}\;
\epsilon_{2,\lambda_2}^{\nu,*} \epsilon_{2,\lambda_2}^{\nu_1}\;
\epsilon_{3,\lambda_3}^{\rho,*} \epsilon_{3,\lambda_3}^{\rho_1}\; {\cal A}^{g}_{\mu_1 \nu_1 \rho_1}(s,t,u, m_t)\,,\\
F^{q}_i(s,t,u,m_t)&=\sum_{pol}\mathcal{P}_{\mu}^{qi} 
\epsilon_{3,\lambda_3}^{\mu *} \epsilon_{3,\lambda_3}^\nu \, \mathcal{A}^{q}_\nu(s,t,u,m_t)\,.
\end{split}
\label{eq:projdef}
\ee
Explicit expressions for projection operators can be found in Refs.~\cite{Melnikov:2016qoc,Melnikov:2017pgf}. 

Both FORM~\cite{Vermaseren:2000nd} and FormCalc~\cite{Hahn:1998yk} have been independently used to implement 
the algebraic manipulations related to the projection in $d$ dimensions. The resulting form factors are 
expressed as linear combinations of scalar integrals
\begin{equation}
\I_{\rm top}(a_1,a_2,...,a_8,a_9)
= \int 
\frac{\mathfrak{D}^dk \mathfrak{D}^dl}{[1]^{a_1} [2]^{a_2}  [3]^{a_3}  [4]^{a_4}  [5]^{a_5}  [6]^{a_6}  [7]^{a_7}  [8]^{a_8}  [9]^{a_9} 
},
\label{eq3.5a}
\end{equation}
where the integration measure is chosen to be 
\be
\mathfrak{D}^dk = (-s)^{(4-d)/2}\frac{(4\pi)^{d/2}}{i\Gamma(1+\epsilon)} \int \frac{d^dk}{(2\pi)^d}.
\ee

\begin{table}[t!]
\begin{center}
\begin{tabular}{| c | l | l | l|}
\hline
Prop. & Topology PL1 & Topology PL2 & Topology NPL \\
\hline
$[1]$ & $ k^2$ &                                  $k^2-m_t^2$  &                         $k^2-m_t^2$ \\
$[2]$ & $(k-p_1)^2$ &                         $(k-p_1)^2-m_t^2$  &               $(k+p_1)^2-m_t^2$ \\
$[3]$ & $(k-p_1-p_2)^2$ &                   $(k-p_1-p_2)^2-m_t^2$  &        $(k-p_2-p_3)^2-m_t^2$ \\
$[4]$ & $(k-p_1-p_2-p_3)^2$ &           $(k-p_1-p_2-p_3)^2-m_t^2$  & $l^2-m_t^2$ \\
$[5]$ & $l^2-m_t^2$ &                         $l^2-m_t^2$  &                         $(l+p_1)^2-m_t^2$ \\
$[6]$ & $(l-p_1)^2-m_t^2$ &                $(l-p_1)^2-m_t^2$  &                $(l-p_3)^2-m_t^2$ \\
$[7]$ & $(l-p_1-p_2)^2-m_t^2$ &         $(l-p_1-p_2)^2-m_t^2$  &        $(k-l)^2$ \\
$[8]$ & $(l-p_1-p_2-p_3)^2-m_t^2$ &  $(l-p_1-p_2-p_3)^2-m_t^2$  & $(k-l-p_2)^2$ \\
$[9]$ & $(k-l)^2-m_t^2$ &                     $(k-l)^2$  &                               $(k-l-p_2-p_3)^2$ \\
\hline 
\end{tabular}
\caption{Feynman propagators of the three integral families, see Eq.~\eqref{eq3.5a}.} \label{tab:topos}
\end{center}
\end{table}
\noindent
Scalar integrals that appear in the form factors  belong to one of the three integral families that 
we refer to as  
$\{\text{PL1},\text{PL2},\text{NPL}\}$. 
Sets  of  propagators that define each topology are shown in Table~\ref{tab:topos}. 

After an amplitude is projected on a form factor, 
all scalar integrals are reduced to a set of master integrals (MI) using 
the integration by parts identities (IBP)~\cite{Tkachov:1981wb,Chetyrkin:1981qh}. 
The reduction has been previously performed in Refs.~\cite{Melnikov:2016qoc,Melnikov:2017pgf} using 
public versions of 
 FIRE5~\cite{Smirnov:2008iw,Smirnov:2014hma}, Reduze2~\cite{Bauer:2000cp,Studerus:2009ye,vonManteuffel:2012np,fermat} 
and an in-house routine written in FORM~\cite{Vermaseren:2000nd}.\footnote{We are 
indebted to L.~Tancredi for his decisive contribution to the reduction to master integrals 
for this problem.} 
The MIs are computed by solving differential equations in kinematic variables; the differential 
equations are solved 
perturbatively, expanding 
in the small parameters $\kappa$ and $\eta$, as will be explained in  Section~\ref{sec:masters}. 

We note that  MIs contain logarithmic singularities $\propto \log( m_t^2) \sim 
\log{(\kappa)}$ as $\kappa \to 0$. These are mass singularities that are expected 
to be present in the high-$p_\perp$ kinematics. 
In addition, there are   Feynman integrals that develop  logarithmic singularities 
$\propto \log{(\eta)} \sim \log(m_h^2)$ as $\eta \to 0$;  
 This happens  whenever all the 
massless external partons couple directly to massless internal propagators, 
such as for example is the case for 
the top center diagram in Figure~\ref{fig::feyndiag2loopgg}. 
The resulting MI which appear after pinching this diagram also 
contains logarithmic singularities 
$\propto \log{(\eta)}$. For these MI 
it is possible to cut massless propagators in their corresponding diagrams such that the squared momentum flowing 
into the cut equals $m_h^2$ and therefore we 
expect a singular behavior as $m_h^2\to0$. Note that 
the top loop itself always gets screened by the top mass and therefore the $\log{(\eta)}$ 
singularities are attributed to a
specific scaling of the loop momenta running through massless propagators. 

Since the Higgs boson always couple to top quarks, we expect that 
all the $\log( m_h^2)$ 
singularities are the artifacts of computational procedure 
and that they should cancel in the final result
for form factor. We have confirmed this expectation by an explicit computation.

Another interesting point is that 
three sectors of non-planar MI (one sector with six propagators and two top sectors with 
seven propagators) have integrals whose expansion around $\kappa \to 0$ 
starts  with {\it non-integer} 
powers of $\kappa$, i.e. $I \sim \kappa^{-1/2} \sim m_t^{-1}$.  
This non-analytic behavior indicates contributions to scalar integrals {\it beyond} 
the standard soft-collinear paradigm. It is interesting to see, however, that 
none of these non-analytic terms survives  in 
final results for physical amplitudes. 

To conclude, after the results for MIs are used to calculate  the  unrenormalized form factors, 
the form factors are written as an expansion in  $\kappa$ and $\eta$
\begin{equation} 
\begin{split} 
\lim_{\frac{m_h}{m_t}  \to 0, m_t  \to 0} (F^i)_j^{(1),\text{un}}(\kappa,\eta,z) &= \kappa\, 
\sum_{n=-2}^0\; \epsilon^n\; \sum_{a=0}^1\, \kappa^a \sum_{b=0}^2\, f_{a,b,i,j}^{(1l,n)}(z)\, \log^b{\kappa}\,, 
 \\
\lim_{\frac{m_h}{m_t}  \to 0, m_t  \to 0} (F^i)_j^{(2),\text{un}}(\kappa,\eta,z) &= \kappa\, \sum_{n=-2}^0\; \epsilon^n\; \sum_{a=0}^1\, \kappa^a \sum_{b=0}^4\, f_{a,b,i,j}^{(2l,n)}(z)\, \log^b{\kappa}. \label{eq:expform2}
\end{split} 
\end{equation} 

The Yukawa coupling and the helicity flip in one of the quark lines contribute each a factor of $m_t$, which results in the 
overall factor of $\kappa$ in the above result. In  Eq.~\eqref{eq:expform2} we retain 
terms that are leading powers  in the 
squared Higgs mass $\eta$ and up to next-to-leading power in the squared top quark mass. 
Since there are no logarithms in $\eta$ in the final result, we could have put 
$\eta\rightarrow 0$ from the beginning. However, 
in our computation we did not do this and kept $m_h^2\sim \eta \ne 0$  
throughout the calculation,\footnote{The planar master integrals corresponding to 
the case where $m_h^2=\eta=0$ have been computed in \cite{Caron-Huot:2014lda,Becchetti:2017abb}.} 
 only cutting off the expansion of form factors  
to leading power after inserting  the MIs. It was argued 
in e.g. Ref.~\cite{Braaten:2017lxx} for the quark channels that an 
expansion to leading power  in $\eta$ gives a good 
approximation to the full amplitude  with non-zero Higgs mass. We  have checked  
this statement explicitly by comparing expanded and un-expanded one-loop amplitudes for both quark 
and gluon channels.  We conclude that   Eq.~\eqref{eq:expform2} 
is expected to provide a reasonable description of the  form factors with non-zero Higgs 
and top mass.   We will next describe how to  compute the MIs  using the method 
of differential equations.
  

\subsection{Solving for the two-loop master integrals} 
\label{sec:masters} \setcounter{equation}{0} 
\numberwithin{equation}{section} 

The master integrals with seven propagators 
correspond to  Feynman diagrams shown in 
Figs.~\ref{fig::feyndiag2loopgg};
all other MIs that contain  six or even less propagators can be obtained from the highest-level ones 
by pinching.  
We note that all the master integrals for $H+$jet production  were recently 
computed  in an approximation $m_{q=b}^2 \ll s \sim t \sim u \sim m_h^2$, 
in Refs.\cite{Melnikov:2016qoc,Melnikov:2017pgf}. In this paper we 
are instead interested in computing master integrals 
 for high energies and transverse momenta   $m_{q=t}^2\ll s,t,u$ 
in a situation when the quark mass is   larger than the Higgs mass, $m_h^2\ll m_{q=t}^2$.

To derive differential equations, we 
start by taking derivatives of the integrals with respect to the kinematic invariants $m_t^2,s,t,u$. The 
derivatives with respect to the Mandelstam variables can be expressed in terms of linear combinations 
of derivatives with respect to the four-momenta of the external particles
\begin{eqnarray}
s\, \partial_{s}&=&\frac{1}{2}\left(p_1 \cdot\partial_{p_1}+p_2\cdot\partial_{p_2}-p_3\cdot\partial_{p_3}\right), \nonumber\\
t\, \partial_{t}&=&\frac{1}{2}\left(p_1\cdot\partial_{p_1}-p_2\cdot\partial_{p_2}+p_3\cdot\partial_{p_3}\right), \label{eq:deqsij}\\
u\, \partial_{u}&=&\frac{1}{2}\left(-p_1\cdot\partial_{p_1}+p_2\cdot\partial_{p_2}+p_3\cdot\partial_{p_3}\right). \nonumber
\end{eqnarray}
Here we use the notation $p_i \cdot \partial_{p_j} = p_{i,\mu} \, \partial/ \partial p_j^\mu$. 
The derivatives with 
respect to dimensionless variables defined  in  Eq.~\eqref{eq:variables} 
are  related to  above differential operators through the following equations 
\begin{gather}
\partial_{\eta}=4s\kappa\partial_{t}, \quad \partial_{\kappa}=s\left(4\eta\partial_{t}-\partial_{m_t^2}\right), \quad \partial_{z} =s\left(\partial_{u}-\partial_{t}\right).
\label{eq:deqxyz}
\end{gather}
We apply the derivatives in Eqs.~(\ref{eq:deqxyz},\ref{eq:deqsij}) 
to the set of master integrals and use integration-by-parts identities to reduce  all the integrals 
back to master integrals.  This procedure  leads to 
a linear system of coupled partial DE for all the MIs that we will  denote in this Section by $\{\I_i\}$. 
After  expressing the MIs in terms of the chosen variables, the derivative with respect to the 
 Mandelstam variable $s$ 
becomes  trivial and provides the mass dimension of the MIs. Therefore, it suffices to solve the MIs for the 
case $s=1$ and re-introduce it  back at the end of the calculation.

The DEs take the following form
\begin{equation}
\partial_k\I_i(\kappa,\eta,z,\epsilon)= \sum_j A^k_{ij}(\kappa,\eta,z,\epsilon)\, \I_j(\kappa,\eta,z,\epsilon), 
\quad k \in \{ \kappa,\, \eta,\, z \}. \label{eq:deqm2yz}
\end{equation}
The matrices $A^{\kappa,\eta,z}$ are sparse and can be put in a triangular form. We may then 
solve the system starting from the simplest integrals, which then serve as inhomogeneous contributions to the 
DEs of integrals with more propagators. The integrals which depend on a  single scale, e.g. the two-loop tadpole 
integrals, are computed independently and serve as an input for the DEs.

The three matrices $A^k$ are rational functions of $\eta, \kappa,z$ and $\epsilon$. The MIs have been chosen such 
that the  dependence on space-time dimensionality  $d$ 
does not mix with the kinematic variables inside the denominators that appear in $A^k$. The matrices 
have singularities at $\eta=0,-1/2,-1$, in other words at $m_h^2=0, m_h^2=2m_t^2$ and $m_h^2=4m_t^2$ respectively. The pole at 
$m_h^2=2m_t^2$ is expected to be spurious and can be avoided by taking canonical combinations of the MI as 
in~\cite{Bonciani:2016qxi}. At the point $\eta=m_h^2=0$ there are singularities at $\kappa=0,-1/4,-(1+z)/4,-z/4$, which 
corresponds to poles at $m_t^2=0 $ and $s,t,u=4m_t^2$ respectively. The 
latter three poles arise from the top threshold when the invariant mass of a pair of final state particles in the processes 
in Eq.~\eqref{eq:Htoqqg} is equal to $2 m_t$. These considerations imply that the matrices can be conveniently 
expanded in $m_h^2/(4m_t^2)=-\eta$ and $4m_t^2/s=-4\kappa$ and the DE then solved perturbatively in small $\eta$ and $\kappa$. 
The order of expanding in $\eta$ and $\kappa$ is irrelevant. Furthermore,  since  the DEs have singularities at 
both $\eta=0$ and $\kappa=0$, the solutions are expected to contain 
terms beyond a usual analytic Taylor expansion in $\eta$ and $\kappa$. 
The structure of the differential equations 
implies the following ansatz
\begin{equation}
\I_i(\kappa,\eta,z,\epsilon)=\sum_{j,k,l,m\in\mathds{Z},n\in\mathds{N}} c_{i,j,k,l,m,n}(z,\epsilon)\, \eta^{j-k\epsilon} \kappa^{l/2-m\epsilon}\, \log^n(\kappa). \label{eq:ansatz1}
\end{equation}

A more detailed analysis of the differential equations shows that at two loops there are at most two powers of 
$\kappa^{-\epsilon}$ and $\log(\kappa)$ and at most one  power of $\eta^{-\epsilon}$. The following simpler ansatz 
therefore suffices
\begin{equation}
\I_i(\kappa,\eta,z,\epsilon)=\sum_{j\geq 0, l\geq -3}\sum_{k=0}^1\sum_{m=0}^2\sum_{n=0}^2 c_{i,j,k,l,m,n}(z,\epsilon)\, \eta^{j-k\epsilon} \kappa^{l/2-m\epsilon}\, \log^n(\kappa). \label{eq:ansatz2}
\end{equation}
The  maximal value for the powers $j,l$ of the variables $\eta$ and $\kappa$,  respectively, 
are chosen such that we can 
expand the form factors to leading power 
 in $\eta$ and to next-to-leading power  in $\kappa$. We note that this requires computing some of the MI to 
higher suppressed powers in $\eta$ and $\kappa$.

As we already alluded to  in  the paragraph 
above Eq.~\eqref{eq:expform2}, we need to include powers of $\eta^{-\epsilon}$ in the ansatz 
for exactly six MI, which all appear in the planar topology PL1. 
A detailed study of these six MI shows that they indeed have terms that scale as 
$\eta^{-\epsilon}$ when $\eta\to 0$ but there are no $1/\eta$ singularities. For all other MI, the expansion in 
$\eta\to 0$ correspond to a simple Taylor expansion. We conclude that none of the MI have singularities 
in $1/\eta$, which fixes the lower bound 
of the index $j$ in the sum of Eq.~\eqref{eq:ansatz2} to zero. On the other hand, the lower bound on the 
index $l=-3$ in the sum of Eq.~\eqref{eq:ansatz2} follows directly from the structure of the DE.
Furthermore the DE allow half-integer powers of $\kappa$ for exactly three non-planar four-point sectors of MIs. 
Finally, all 
the terms that are non-analytic in $\eta$, i.e. contain factors of  
$\eta^{-\epsilon}$, or contain half-integer powers of $\kappa$,  cancel when final 
form factors are computed.  However, we keep them in our ansatz 
and compute them when solving for master integrals since 
their cancellation provides a good check of the correctness of the 
calculation.

The coefficient functions $c_{i,j,k,l,m,n}$ 
depend on $z$ and $\epsilon$. We determine them by substituting  
the  ansatz for integrals in Eq.(\ref{eq:ansatz2})
into the  differential equations and equating 
terms  with the same powers of $\eta,\kappa$ and $\log(\kappa)$
on both sides of the relevant equations. This procedure relates 
the $c_{i,j,k,l,m,n}$ coefficients 
to each other via a system of linear {\it algebraic} equations. We note that the DEs 
allow powers 
of $\eta^{-1}$ in our complete ansatz in Eq.~\eqref{eq:ansatz1}. Therefore,  {\it requiring} that 
solutions to DEs do not 
contain  poles in $\eta$  provides additional relations between coefficient functions. 

Some of the coefficient functions $c_{i,j,k,l,m,n}$ remain undetermined after 
solving  the differential equations  in $\eta$ and $\kappa$. 
However, we can solve the DEs in such a way that these undetermined coefficient functions appear in the 
leading power expansion of $\eta$, i.e. in terms that correspond  to 
$j=0$ in our ansatz  Eq.~\eqref{eq:ansatz2}. 
The ``massless'' coefficients $c_{i,0,0,0,0,0}$ correspond to 
a completely massless ``version'' of the MIs which is obtained by setting  $m_h^2$ and $m_t^2$  to zero 
at the integrand level. These integrals are well-known and serve as an input 
in our calculation. Indeed, all the needed planar massless master integrals have been computed 
in Refs.~\cite{Smirnov:1999wz,Henn:2013pwa}.\footnote{Note that in the massless limit, 
the integral $I^{PL2}_{1,1,1,-1,1,0,1,1,1}$ can be written as 
$\frac{1}{\epsilon}(I^{PL2}_{1,1,1,0,1,0,1,1,1}-u I^{PL2}_{1,1,1,0,1,0,1,1,2})$ plus lower sub-topologies. 
The order $\epsilon$ pieces of the two planar master integrals are of  weight five, but 
in the difference only terms with weight four survive. These terms  were computed 
in Ref.~\cite{Gehrmann:2000xj}.} 
The non-planar massless master integrals have been taken 
from Refs.~\cite{Tausk:1999vh,Anastasiou:2000mf,Argeri:2014qva}.

After fixing the massless coefficients to the known computed massless MI, we are left with undetermined 
coefficients $c_{i,0,k,l,m,n}$. To find them,  we use the ansatz in Eq.~\eqref{eq:ansatz2} in the 
$z$ differential 
equation and again  equate  terms with matching powers of $\eta,\kappa$ and $\log(\kappa)$ 
on both sides of the differential 
equation. The  DEs in $z$ are relatively simple and can be solved order by order in an expansion 
in $\epsilon$. 
Similar to the  case of  massless MIs, 
the solutions are expressed in terms 
of Harmonic Polylogarithms (HPLs)  which form  a subset of the  Goncharov polylogarithms
\begin{gather}
\text{G}(\underbrace{l_1,\cdots ,l_n}_{\text{weight n}};z):=\int_0^z dz' \frac{\text{G}(l_2,\cdots ,l_n;z')}{z'-l_1}, \nonumber\\
 \quad \text{G}(;z)=1, \ \ \ \ \text{G}(\underbrace{0,\cdots ,0}_{\text{n times}};z)=\frac{1}{n!}\log^n(z).\label{eq:2dHPL}
\end{gather}

The letters that we encounter in the $z$ DEs  are very simple; the  alphabet reads
\begin{equation}
l_i\in \{0,\, -1\}. \label{eq:letters}
\end{equation}
The first letter corresponds to branch points at $s=0$ or $u=0$ when $\eta=m_h^2=0$, the second to $t=0$. After 
solving the equations in $z$, we 
expand the solutions in $\epsilon$ keeping all the terms 
up to weight four\footnote{Some coefficients are pure rational functions in $z$ after expanding 
in $\epsilon$. In these cases we expand to exactly four orders higher in $\epsilon$, starting from the highest pole in 
$\epsilon$ of that coefficient.}  
\begin{equation}
c_{i,j,k,l,m,n}(z,\epsilon)= \sum_{r=r_0^{(i,j,k,l,m,n)}}^{r_0^{(i,j,k,l,m,n)}+4}\epsilon^r\, c_{i,j,k,l,m,n}^{(r)}(z). \label{eq:mastersol}
\end{equation}
The powers of $\epsilon$ in the expansion are bounded below by $r = -4$. 
Typically, individual coefficient functions have 
higher singularities in $\epsilon$ than the expanded solution.  This feature is understandable since 
massive internal particles screen infra-red and collinear singularities; for this reason, 
full results for master integrals should typically be less singular in the $\ep \to 0$ limit than 
their  massless branches. 

After solving the DEs in $z$ we are left with unknown integration constants 
that need to be determined. For the  MIs in the planar topologies PL1 and PL2 we could 
fix many of the constants by requiring that the unphysical singularities at $z=-1$ cancel. 
We are allowed to do this 
since the corresponding planar diagrams do not have any cuts in the $t$-channel, but only in 
$s$ and $u$. After requiring that these unphysical branch points at $t=0$
vanish,  all of the constants in topology PL2 become  fixed. We are 
left with one constant in the family PL1 and six in the family NPL that we need 
to determine in some other   way. In the next Section we will explain how we computed 
these constants using the Mellin-Barnes representation of the relevant integrals.  

We note that in order to compute the amplitude to order ${\cal O}(\epsilon^0)$, we are required to compute 
coefficient functions of some integrals to weight {\it five} and a few even to weight {\it six}. 
By using the DEs in 
$\eta$ and $\kappa$, we could find many connections between contributions 
of weights five and six to the  coefficient functions of the MIs. 
After substituting MIs  into the amplitude, most of the 
unknown weight five and all of the weight six contributions  
cancel amongst each other. The few weight five pieces that 
 are left, appear only in the planar families PL1 and PL2 and for these 
we needed to integrate the DEs in $z$ to weight five. However the 
integration constants of these  weight-five 
contributions   {\it cancel} 
in the final result for  the amplitude and therefore they did not have to be computed. 

The expansions  of the  MIs
in $\kappa$ and $\epsilon$  have been, whenever possible, numerically compared with 
FIESTA~\cite{Smirnov:2015mct} at the point $m_h^2=\eta=0$ and an 
agreement was found within the integration errors of FIESTA. We include with this paper ancillary files 
that contain our solutions for all MIs in the form of the ansatz in Eq.~\eqref{eq:ansatz2}, expanded in 
$\eta$ and $\kappa$ to orders that are sufficient  
to compute the amplitude to leading  and next-to-leading power  in $\eta$ and $\kappa$,  respectively.

\subsection{Integration  constants and numerical checks using  Mellin-Barnes} 
\label{sec:bound}
\numberwithin{equation}{subsection}

As we already mentioned several times, by solving  differential equations we determine 
master integrals up to integration constants  that have to be determined in a different  way. 
Many integration  constants can be  fixed by requiring that integrals have regular limits at certain 
singular points of the differential equations, for example at $\eta \to 0$. 
However, there are seven integration constants 
that are  left undetermined by these considerations and we have to compute them explicitly. To accomplish 
that, we use the 
Mellin-Barnes representation to calculate the  relevant  master integrals  
at certain  kinematic points  and then match the results to solutions of  
differential equations. 

The Mellin-Barnes representation has been used before to compute the massless coefficient functions of some 
planar~\cite{Tausk:1999vh} and non-planar 
master integrals~\cite{Anastasiou:2000kp}. 
Since we  relate the coefficient functions corresponding to higher powers 
in  expansion in $\eta$  to coefficient functions at  leading power in the $\eta$ expansion, 
all undetermined integration constants 
appear in the coefficient functions that can be computed by  setting $\eta$ to zero. 
In other words we have  to keep the non-vanishing 
top quark  mass\footnote{As we mentioned, the massless coefficients are already computed and 
therefore the unknown integration constants always multiply $\kappa^{l-m\epsilon}$ with either $l$ or $m$ non-zero
in the ansatz.} but we may set $m_h^2=0$ in our Mellin-Barnes computation from the very beginning.  
The Mellin-Barnes representation is ideal for organizing the computation as an expansion in a small 
parameter and isolating different $\kappa$-branches since different powers of $\kappa$  appear naturally 
after  residues are computed in Mellin-Barnes integrals. 


\begin{figure}[t!]
\centering
\includegraphics[width=0.25 \linewidth]{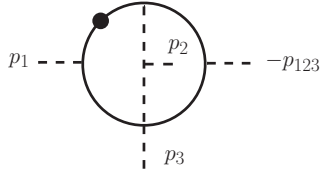}
\caption{The two-loop scalar Feynman corresponding to $\I_{\rm NPL}(0,1,1,1,2,0,1,1,0)$.}
\label{fig::feyndiagNPLt6}
\end{figure} 

We will  consider the integral $\I_{\rm NPL}(0,1,1,1,2,0,1,1,0)$, shown in Fig.~\ref{fig::feyndiagNPLt6}. It 
reads 
\begin{eqnarray}
\I^{\rm top}_{011120110}
&= &\int 
\frac{\mathfrak{D}^dk \mathfrak{D}^dl}{((k_1+p_1)^2-m_t^2)((k_1-p_{23})^2-m_t^2)(k_2^2-m_t^2)((k_2+p_1)^2-m_t^2)} \nonumber\\
&&\times \frac{1}{((k_1-k_2)^2)^{1+\delta}((k_1-k_2-p_{23})^2)^{1-\delta}}. \label{eq:MB0}
\end{eqnarray}
Note that we introduced additional parameters $\delta$ to define $\I^{\rm top}_{011120110}$; we will explain 
below why this is required. 

We are  interested in computing an integration constant 
of the coefficient function that corresponds to  a factor $\kappa^{-1-2\epsilon}$.  
We choose the kinematic point 
$s=u=-1, t=2, m_h^2=0$. 
The integral is well defined and regulated by dimensional 
regularization. However, the region integrals that represent the 
term $\propto\kappa^{-1-2\epsilon}$ are not regulated by the dimensional parameter 
$\epsilon$.  This can be already seen 
from the solution of the DE that  predicts terms $\propto\kappa^{-1-2\epsilon}\log^{1,2}(\kappa)$.
Such   non-analytic behavior is typically cured by an introduction of analytic 
regulators in the context of asymptotic expansion 
of Feynman integrals \cite{Smirnov:1997gx,Jantzen:2011nz}.  The parameter 
$\delta$ introduced in Eq.~\eqref{eq:MB0} is the analytic regulator that 
makes expansion of all branches the integral $\I^{\rm top}_{011120110}$ well-defined. 

To proceed further,  we introduce Feynman parameters and integrate over  two loop momenta. The 
integral can be expressed as powers of the Symanzik $\mathcal{U}$ and $\mathcal{F}$ polynomials
\begin{gather}
-\int_0^{\infty} \left(\prod_{i=1}^6 dx_i\right)\frac{\delta(1-\sum_{I} x_i)\Gamma (2 \epsilon+3) x_4 x_5^{-\delta } x_6^{\delta }}{\Gamma (1-\delta ) \Gamma (\delta +1) \Gamma (\epsilon+1)^2} \Bigg [
(x_3+x_4) (x_5+x_6)+x_1 (x_3+x_4+x_5+x_6) \nonumber\\
 +x_2 (x_3+x_4+x_5+x_6) \Bigg ]^{3 \epsilon+1}  \; \Bigg [ 
x_2 x_3 x_5+x_1 x_3 x_6-2 x_2 x_4 x_6 +\kappa (x_1+x_2+x_3+x_4) 
 \nonumber\\
\times ((x_3+x_4) (x_5+x_6)+x_1 (x_3+x_4+x_5+x_6)+x_2 (x_3+x_4+x_5+x_6)) -i0 \Bigg ]^{-2 \epsilon-3}.\nonumber
\end{gather}

The sum inside the delta function can be chosen to be any combination of the Feynman 
parameters \cite{Cheng:1987ga,Bjoerkevoll:1992cu}. We have chosen the 
delta function as $\delta(1-x_1-x_2)$. The integration over the Feynman parameters are nontrivial but may be performed by 
using the method of Mellin-Barnes. Namely,  we may split up terms inside the brackets by introducing 
Mellin-Barnes integration parameters
\begin{gather}
\frac{1}{(x+y)^\lambda}=\frac{1}{2\pi i}\int \limits_{-i\infty}^{+i\infty} dz \frac{y^z}{x^{z+\lambda}}\frac{\Gamma(-z)\Gamma(\lambda+z)}{\Gamma(\lambda)}.
\label{eq:MB1}
\end{gather}

The contour runs parallel to the imaginary axis in the complex $z$-plane and is chosen such that 
the singularities of $\Gamma(-z)$ and $\Gamma(\lambda+z)$ are  to the right (left),  
respectively of the integration contour. 
After we integrate over  Feynman parameters,  we are left with the following Mellin-Barnes integral to perform
\begin{equation} 
\begin{split} 
& \I^{\rm top}_{011120110}=-\int \limits_{-i\infty}^{+i\infty} \left(\prod_{i=1}^4 dz_i\right)(-2-i 0)^{-2 \epsilon-z_1-z_2-z_3-3}\kappa^{z_1} \Gamma (-z_1) \Gamma (-z_2) \Gamma (z_2+1) \Gamma (-z_3) 
\\
&\times\, \frac{\Gamma (-z_4) \Gamma (-\epsilon-z_1-1) \Gamma (z_4-\epsilon) \Gamma (z_3-\delta +1) \Gamma (-2 \epsilon-z_1-z_2-2) \Gamma (z_2+z_3+z_4+1)}{\Gamma (1-\delta ) \Gamma (\delta +1) \Gamma (\epsilon+1)^2 \Gamma (-2 \epsilon-2 z_1-1) \Gamma (-3 \epsilon-z_1-1) \Gamma (-2 \epsilon-z_1-1)} 
\\
&\times\, \Gamma (2 \epsilon+z_1+z_2+z_3+3) \Gamma (-2 \epsilon-z_1-z_3+\delta -2) 
\Gamma (-\epsilon-z_1-z_2-z_3-z_4-1). 
\label{eq:MB2}
\end{split}
\end{equation}

The Mellin-Barnes integrations can be performed with the help of packages collectively known 
as MBTools \cite{MBTools}.  For example, the 
contours of the $z_{1..4}$ integrals  can be systematically 
deformed ~\cite{Czakon:2005rk} in a way that allows one  to take the limit $\delta \to 0$. 
Indeed, because $\delta$ is an analytic regulator, we need to take $\delta \to 0$ at fixed $\epsilon$ 
and then deform the contour further to extract poles in $\ep$ and, eventually, arrive at the $\epsilon$ 
expansion. We note that, as follows from Eq.~\eqref{eq:MB2}, poles in $z_1$ correspond  to different powers 
of $\kappa$; for our purposes we require the pole at $z_1 = -1 - 2\epsilon$. 
After extracting  the $\kappa^{-1-2\epsilon}$ branch 
and expanding  the result in 
$\epsilon$,  we use the Barnes lemma to perform the Mellin-Barnes integrations.  In most cases, 
these integrations are straightforward. However, we also obtain a contribution which requires 
to deal with the integrand  that contain polygamma functions. A typical integral  reads
\begin{eqnarray}
 I = \int_{0^--i\infty}^{0^-+i\infty} dz_4 && \Gamma (2-z_4) \Gamma (z_4-1) \Gamma (-z_4) \Gamma (z_4)
\Big [ 2 (\psi ^{(0)}(2-z_4))^2+(\psi ^{(0)}(z_4))^2  \nonumber\\
&&  +2 \psi ^{(1)}(2-z_4)+\psi ^{(1)}(z_4) \Big ].
\label{eq:MB4}
\end{eqnarray}
The integration over $z_4$ is performed using  
the method of residues. The integration contour runs along the imaginary axis 
with $\text{Re}(z_4)$ small and negative. We may close the integration contour 
to the left as the integrand will vanish fast enough along the half circle in the left complex plane with infinite radius. 
By Cauchy's theorem 
we pick up the ladder of residues in $z_4$ in the left complex plane, $\text{Re}(z_4)<0$. 
Application of this procedure leads to the following representation 
\begin{gather}
I = - \frac{1}{3} \sum_{n=1}^{\infty}\frac{1}{n^4 (n+1)^2} \left(\left(\pi ^2 n^2+6\right) (n+1)^2+3 n \left(2 \left(3 n^2-3 \left(n^2+n\right)^2 \psi ^{(1)}(n+1)-1\right) \psi ^{(0)}(n+1) \right.\right. \nonumber\\
\left.\left. +3 n (n+1)^2 \psi ^{(0)}(n+1)^2-n (n+1) ((n-3) \psi ^{(1)}(n+1)+2 n (n+1) \psi ^{(2)}(n+1))\right)\right).  \nonumber
\end{gather}
These sums can be performed with e.g.  the  XSummer~\cite{Moch:2005uc}.

The final result for the  ${\cal O}(\kappa^{-1-2\ep})$ branch 
of the  $\I^{\rm top}_{011120110}$ at the kinematic point 
$s=u=-1, t=2, m_h^2=0$   reads 
\be
\label{eq:MB3}
\begin{split} 
 \I^{\rm top}_{011120110} & \sim  
\kappa^{-1-2\epsilon}  \left\{\frac{\log ^2(\kappa)}{4 \epsilon}+\log (\kappa) \left(\frac{1}{\epsilon^2}-\frac{3 \epsilon \zeta (3)}{2}+\frac{\log (2)-i \pi }{2\epsilon}-\frac{\pi ^2}{12}\right)
 \right. 
\\
&  \left. 
+\frac{1}{\epsilon^3}+\frac{\log (2)-i \pi }{\epsilon^2}
+\frac{-\frac{1}{6} \pi  (\pi -6 i)+\frac{\log ^2(2)}{4}
+\left(-1-\frac{i \pi }{2}\right) \log (2)}{\epsilon}
 \right. \\
&  \left. 
+\frac{1}{12} \left(i \pi ^3-18 \zeta (3)\right)-\frac{\pi ^2 \log (2)}{12} 
+\epsilon \left(\frac{3 i \pi  \zeta (3)}{2}
-\frac{3}{2}  \zeta (3) \log (2) -\frac{7 \pi ^4}{240 }\right)\right\}. 
\end{split}
\ee
We then match the solution of the differential equation to this result and determine the integration 
constant. 


In addition to determination of constants, 
we also used the Mellin-Barnes representation 
for  numerical checks of our solutions for master integrals. 
Namely for non-planar MIs that, in the $\kappa \to 0$ limit develop power-like singularities with 
half-integer exponents, 
we were unable to use FIESTA for numerical checks. 
For such integrals  we compared {\it individual} coefficient functions in $\kappa$ 
with the  Mellin-Barnes representation and found perfect 
agreement for all of them. In particular, the massless contributions as well as other 
coefficient functions that are completely fixed by the DE, all agree with the Mellin-Barnes result. 
Note that this is a nontrivial check on both the  solution for the differential 
equation  and the  Mellin-Barnes 
representation  that we used to extract integration constants for certain branches. 
One example  of the coefficient functions that are completely fixed by the DEs are those 
corresponding to the $\kappa^{-1-2\epsilon}\log(\kappa)$ and $\kappa^{-1-2\epsilon}\log^2(\kappa)$ 
terms that appear in our solution of the 
above integral $\I^{\rm top}_{011120110}$, which we have checked to agree exactly with the corresponding 
logarithms in $\kappa$ 
in Eq.~\eqref{eq:MB3} for the chosen  kinematic point. 

\section{Helicity amplitudes} 
\label{sec:helamp} \setcounter{equation}{0} 
\numberwithin{equation}{section}

Once the master integrals are computed, we use them 
to derive the form factors and calculate  the analytic 
expressions for helicity amplitudes. 
We define positive and negative helicity spinors for massless external quarks and gluons 
in the standard  way  (see e.g.~\cite{Dixon:1996wi})
\begin{align}
&\epsilon_{i,+}^{\mu}(p_i) = \frac{\langle q | \gamma^\mu | i ]}{\sqrt{2} \langle q \, i \rangle }
\,,
\quad \,\,\,\,\,\, \epsilon_{i,-}^{\mu}(p_i) = - \frac{[ q | \gamma^\mu | i \rangle}{\sqrt{2} [ q \, i ] }\,,
\\
&u_{+}(p) = v_{-}(p) = | p \rangle \,, \quad  u_{-}(p) = v_{+}(p) = | p ]\,,\nonumber \\
&\bar{u}_{+}(p) = \bar{v}_{-}(p) = [ p | \,, \quad \, \bar{u}_{-}(p) = \bar{v}_{+}(p) = \langle p |\,.
\label{eq:polquark}
\end{align}
Here $q$  is an arbitrary  light-like  reference vector.  For our computation, the reference 
vectors are fixed by gauge conditions outlined in Eq.(\ref{eq:gauge}).

The helicity amplitudes are defined as
\begin{align}
{\cal A}^{g}_{\lambda_1 \lambda_2 \lambda_3}(s,t,u,m_t) &= 
\epsilon_{1,\lambda_1}^\mu (p_1) 
\epsilon_{2,\lambda_2}^\nu (p_2) \epsilon_{3,\lambda_3}^\rho (p_3)
{\cal A}^{g}_{\mu \nu \rho}(s,t,u,m_t), \\
\mathcal{A}^{q}_{\lambda_1 \lambda_2 \lambda_3}(s,t,u,m_t) &= 
\epsilon_{3,\lambda_3}^\mu (p_3)  
\bar{u}_{\lambda_1}(p_1)\, \mathcal{A}^{q}_\mu(s,t,u,m_t)\, v_{\lambda_2}(p_2)\,.
\label{eq:hela}
\end{align}
Eight helicity configurations are needed to describe 
the $H\to ggg$ amplitude. However,  only two of them are independent 
since the other six may be related to them by the use of charge and parity conjugation. 
For the $H\to q\bar{q}g$ amplitude there are  
four possible helicity configurations in total, 
since QCD interactions cannot change the helicity of the massless quarks and 
therefore the helicity of the outgoing quark must be opposite to that of the 
outgoing anti-quark in Eq.~\eqref{eq:hela}.
We have chosen to treat the following amplitudes as independent
\begin{align}
{\cal A}^{g}_{+++}(s,t,u,m_t) &= 
\frac{s}{\sqrt{2} \langle 12 \rangle \langle 23 \rangle \langle 31 \rangle}\; 
\, \Omega^{g}_{+++}(s,t,u,m_t)\,,
\\
{\cal A}^{g}_{+-+}(s,t,u,m_t) &= \frac{[13]^3}{\sqrt{2}\, [12]\, [32]\, s}\; 
\, \Omega^{g}_{+-+}(s,t,u,m_t)\,,
\\
\mathcal{A}^{q}_{-++}(s,t,u,m_t) &= \frac{1}{\sqrt{2}} \frac{[23]^2}{[12]\, s}\, \Omega^{q}_{-++}(s,t,u,m_t).
\label{eq:helampl}
\end{align}

The amplitudes are dimensionless and the helicity coefficients $\Omega^{i}(s,t,u,m_t)$ have a mass 
dimension one. We 
may obtain the other helicity assignments for the amplitude by complex conjugation and by permuting the external legs as follows
\begin{align}
\mathcal{A}^{g}_{++-}(p_1,p_2,p_3) & = \mathcal{A}^{g}_{+-+}(p_1,p_3,p_2)\,, \\
\mathcal{A}^{g}_{+--}(p_1,p_2,p_3) & = \left[ \mathcal{A}^{g}_{+-+}(p_2,p_1,p_3)\right]^*\,, \\
\mathcal{A}^{q}_{+-+}(p_1,p_2,p_3) & = \mathcal{A}^{q}_{-++}(p_2,p_1,p_3)\,, \\
\mathcal{A}^{i}_{\lambda_1\lambda_2\lambda_3}(p_1,p_2,p_3) & = \left[ \mathcal{A}^{i}_{(-\lambda_1)(-\lambda_2)(-\lambda_3)}(p_1,p_2,p_3)\right]^*\,.
\label{eq:allhelampl}
\end{align}
The complex conjugation should only be applied to spinor-helicity structures and not to the helicity 
coefficients $\Omega^{i}(s,t,u,m_t)$. The helicity coefficients can be expressed in terms of the form factors 
introduced in Eq.(\ref{eq:tensor}) as follows 
\begin{equation}
\Omega^{g}_{+++} = 
u
\left( 
 F^g_1 
+ \frac{t}{u} F^g_2 + \frac{t}{s} F^g_3 + \frac{t}{2} F^g_4 
\right ),
\;\;\;
\Omega^{g}_{+-+} = 
\frac{-s^2}{t} \left ( F^g_2 + \frac{u}{2} F^g_4  \right ),
\;\;\;
\Omega^{q}_{-++} = s^2\, F^q_1\,.
\label{eq:Ftoomega}
\end{equation}

We expand the helicity coefficients in the strong coupling constant and extract an overall coefficient $m_t^2/v$ 
in order to have dimensionless one- and two-loop helicity coefficients
\begin{align}
\Omega^{i} = \frac{m_t^2}{v} \, \sqrt{\frac{\alpha_s^3}{\pi}} \left[ \Omega^{i,(1l)}
+ \frac{\alpha_s}{2 \pi} \Omega^{i,(2l)} + \mathcal{O}(\alpha_s^2)\right]\,.
\end{align}
Once the form factors have been renormalized and IR-subtracted, the resulting helicity coefficients will also be 
finite as seen from Eq.~\eqref{eq:Ftoomega}. We are interested in a kinematic region where all Mandelstam 
variables are much larger than the 
top mass $m_t$. Therefore we prefer to define the amplitude in terms of a strong coupling constant that 
runs with $N_f + 1$ active flavors. The relation between the coupling constants defined in the two schemes reads
\begin{equation}
\alpha_s^{(N_f)}(\mu) = \alpha_s^{(N_f+1)} (\mu) \left[ 1 - \frac{\alpha_s^{(N_f+1)}}{6\, \pi}
\log{\left( \frac{\mu^2}{m_t^2}\right)} + \mathcal{O}(\alpha_s^2)\right] \label{eq:NfNfp1}.
\end{equation}
This change in the strong coupling constant leaves the one-loop coefficients unchanged, but the two-loop finite 
remainder of the helicity amplitude changes as follows according to Eq.~\eqref{eq:NfNfp1}
\begin{align}
\overline{\Omega}^{(1l), \fin} = \Omega^{(1l), \fin} \,, \qquad
\overline{\Omega}^{(2l), \fin} = \Omega^{(2l), \fin} - 
\frac{1}{2} \log{\left( \frac{\mu^2}{m_t^2}\right)}\, \overline{\Omega}^{(1l),\fin}.
\label{eq:chscheme}
\end{align}
The helicity coefficients $\overline{\Omega}$ correspond to using a strong coupling constant 
$\alpha_s^{(N_f+1)}(\mu^2)$
that evolves with $N_f + 1$ active flavors.

Unfortunately, analytic  results for helicity amplitudes are too long to be presented here. Instead, 
we provide  ancillary  files that contain finite remainders of the relevant 
helicity amplitudes $\overline{\Omega}$ defined in Catani's subtraction scheme Eq.~\eqref{eq:IRsub} 
with the submission of this paper. 

\subsection{Analytic continuation} 
\label{sec:cont}
\numberwithin{equation}{subsection}

Our goal is to  compute the two-loop amplitudes that are needed to describe  production 
of the Higgs boson with high transverse momentum  at the LHC. 
The relevant production channels are $gg\to Hg, q \bar q \to H g$, $q g \to H q$ and 
$\bar q g \to H  \bar q$. Amplitudes for these  processes are obtained from amplitudes for 
$H \to ggg$ and $H \to q \bar q g$ processes  that we have computed and crossing some final state particles to 
initial states. Our $H \to ggg$ and $H \to q \bar q g$ amplitudes have been computed in the region $t>0,s,u<0$, 
while the physical scattering processes are defined in the kinematic regions where the invariant 
mass of the two initial partons is positive instead. For this reason we are interested in computing the 
amplitudes in the regions $s>0,t,u<0$ and $u>0,s,t<0$ as well, which we will refer to as $(2a)_+$ and $(4a)_+$ respectively. 
The amplitude in these regions can be found by analytically continuing our result from the region $t>0,s,u<0$ which 
we refer to as $(3a)_+$. The three scattering regions are defined in terms of the Mandelstam invariants as
\begin{align}
(2a)_+ \; :& \quad s >0\,, \quad t,u < 0 \,,\\
(3a)_+ \; :& \quad t >0\,, \quad s,u < 0 \,,\\
(4a)_+ \; :& \quad u >0\,, \quad s,t < 0\,.
\end{align}
The above regions correspond to the following physical production channels
\begin{align}
\mbox{region} (2a)_+ &:\, g(-p_1) + g(-p_2) \to H(-p_4) + g(p_3), \, q(-p_2) + \bar{q}(-p_1) \to H(-p_4) + g(p_3), \nonumber\\
\mbox{region} (3a)_+ &:\, g(-p_1) + g(-p_3) \to H(-p_4) + g(p_2), \, \bar{q}(-p_1) + g(-p_3) \to H(-p_4) + \bar{q}(p_2), \nonumber\\
\mbox{region} (4a)_+ &:\, g(-p_2) + g(-p_3) \to H(-p_4) + g(p_1), \, q(-p_2) + g(-p_3) \to H(-p_4) + q(p_1). \nonumber
\end{align}

In the three regions the positive Mandelstam variable receives 
an infinitesimal positive imaginary part
\begin{align}
(2a)_+ \; :& \quad s \to s + i\, 0 \,,\\
(3a)_+ \; :& \quad t \to t + i\, 0 \,,\\
(4a)_+ \; :& \quad u \to u + i\, 0 \,.
\end{align}

The method to perform the analytic continuation from the region $(3a)_+$, where 
our computation has been performed,  to the other two regions 
was explained in Ref.~\cite{Anastasiou:2000mf} and we refer to this  paper for  details. The spinor 
products are left unchanged during the analytic continuation but Harmonic Polylogarithms may 
receive imaginary parts  when continued 
to regions $(2a)_+$ and $(4a)_+$. We introduce the variable $u_j$ for the three scattering regions
\begin{align}
(2a)_+ \; :&\quad u_{2a} = -\frac{u}{s} = - z\,,
\label{eq:ancontchv1}
\\
(3a)_+ \; :& \quad u_{3a} = -\frac{s}{t} = \frac{1}{1+z}\,,
\label{eq:ancontchv2}
\\
(4a)_+ \; :& \quad u_{4a} = -\frac{s}{u} = - \frac{1}{z}\,.
\label{eq:ancontchv3}
\end{align}
Our helicity amplitudes are expressed in terms of the new variables $0 \leq u_j \leq 1$ in the three corresponding regions. 
In this way the imaginary part of the amplitudes is explicit and all the HPL that appear in the results are 
real-valued with the alphabet $\{ 0,1 \}$ in each of the 
 scattering regions. The Harmonic Polylogarithms  can 
be numerical evaluated with the Mathematica package 
HPL \cite{Maitre:2005uu} or the Fortran code CHAPLIN \cite{Buehler:2011ev}. The helicity amplitudes 
$\Omega^q_{+++},\Omega^g_{+-+}\Omega^q_{-++}$ in all three scattering regions are provided in the ancillary 
file  together 
with the submission of this paper.

\section{Conclusions}  
\label{sec:conclusions} \setcounter{equation}{0} 
\numberwithin{equation}{section} 

We computed the two-loop helicity amplitudes that are needed to describe  production 
of the Higgs boson with large transverse momentum at the LHC. The Higgs 
boson interaction with gluons and massless quarks is mediated by loops of massive top quarks.
However, the top quark  mass is considered to be {\it small} compared to Higgs bosons transverse momentum.
Clearly, in this kinematic regime  the Higgs boson mass is also small compared 
to its transverse momentum and we effectively neglect it in our computation. 

Although the dependence of the scattering amplitudes on the Higgs boson mass is simple and 
can be obtained by a simple Taylor expansion, the expansion of the amplitudes 
in the top quark mass  contains non-analytic terms ${\cal O}(\ln(m_t^2/p_\perp^2))$ and is, therefore, 
non-trivial. We construct the expansion of the amplitudes using differential equations for 
master integrals that allow us to obtain both analytic and non-analytic terms in an expansion 
in a controlled way.   Our final results for the amplitudes  are expanded 
to leading power in the Higgs boson mass which, essentially, corresponds to setting the Higgs 
boson mass to zero, and to  next-to-leading power in the top quark mass squared. 
We expect that  the two-loop amplitudes computed in this paper will allow for a robust  
estimate of the number of  Higgs bosons that are expected to be produced at the LHC with very large 
transverse momentum,  and the comparison of this prediction with the 
experimental result~\cite{CMS:2017cbv}. 



\section*{Acknowledgments}

We would like to thank Lorenzo Tancredi and Fabrizio Caola 
for many helpful discussions throughout the project. We 
acknowledge the help of Viktor Papara, Vladimir Smirnov and Bas Tausk with collecting some of the 
completely massless master integrals. The research of K.M. was supported by 
the German Federal Ministry for Education $\&$ Research (BMBF) under grant O5H15VKCCA.
The research of K.K. is supported by the DFG-funded Doctoral School KSETA (Karlsruhe 
School of Elementary Particle and Astroparticle Physics). 

\appendix

\bibliographystyle{bibliostyle}   
\bibliography{Biblio}
\end{document}